\documentclass[%
 reprint,
 amsmath,amssymb,
 aps,
]{revtex4-2}

\usepackage{graphicx}
\usepackage{dcolumn}
\usepackage{bm}
\usepackage{soul}
\usepackage{comment}
\usepackage{xcolor}
\usepackage{bigints}
\usepackage{nicematrix}
\usepackage{physics}
\usepackage{hyperref}
\hypersetup{
    colorlinks=true,
    citecolor=blue,
    linkcolor=blue,
    urlcolor=blue
    }

\begin{document}

\preprint{APS/123-QED}

\title{Holographic superfluid sound modes with bulk acoustic black hole}

\author{Joseph Carlo U.~Candare}
\email{jucandare@up.edu.ph}
\author{Kristian Hauser A.~Villegas}%
 \email{kavillegas1@up.edu.ph}
\affiliation{%
National Institute of Physics, University of the Philippines Diliman
}%

\date{\today}

\begin{abstract}
 
 The sound modes of a flowing superfluid is described by the massless Klein-Gordon equation in an effective background metric. This effective background metric can be designed to mimick a black hole using the acoustic horizon. In this work, we study the AdS/CFT dual of the sound modes in the presence of an acoustic horizon in the bulk. Focusing on fluids with a purely radial flow, we derive the metric tensor for the effective acoustic spacetime and deduce a necessary condition for an acoustic black hole geometry to exist within the fluid. Using specific examples of superfluid velocity profiles, we obtained the source, operator expectation value, Green's function, and spectral density of the dual field theory by solving for the asymptotic behavior of the sound modes near the AdS boundary. In all our examples, the sound modes remain gapless but the excitations are described by branch cuts, instead of poles, which is typical of strongly coupled systems. Furthermore, we calculate the effective Hawking temperature of the dual field theory associated with the bulk acoustic horizon. Lastly, we investigate the near horizon properties and derive the superfluid velocity profile that can give rise to an infrared emergent quantum criticality.
\end{abstract}

\maketitle


\section{Introduction}\label{sec:intro}

The physics of the anti-de Sitter spacetime has gained a lot of interest not only within the gravitational physics community but also in the condensed matter and hadron physics communities due to the anti-de Sitter/Conformal Field Theory (AdS/CFT) duality \cite{zaanen2015holographic,erdmenger2015gauge}. This powerful theoretical framework maps a strongly-coupled quantum many-body system into a weakly-coupled classical field in an asymptotically AdS spacetime background \cite{zaanen2015holographic, erdmenger2015gauge}. It allows for the analytical investigation of non-perturbative systems such as quark-gluon plasma \cite{rosen2009unraveling, sadeghi2013application}, strange metals \cite{sachdev2010strange, PhysRevB.109.085119,PhysRevB.109.155140}, and holographic superconductors \cite{PhysRevLett.101.031601,herzog2009lectures}. 

In this duality, the black hole horizon in the bulk, being thermodynamic and dissipative, plays an essential role as it provides the temperature of the dual field theory via the Hawking temperature. In extremal Reissner-Nordstr\"{o}m blackholes, the near-horizon geometry leads to an emergent local quantum criticality, which might be relevant to strongly correlated systems such as in cuprates \cite{zaanen2015holographic}. 

To realize the holographic superconductors, a complex scalar field is added in the bulk curved spacetime.  The presence of the \lq\lq scalar hair\rq\rq  in the bulk is then dual to the non-zero order parameter describing a superfluid \cite{zaanen2015holographic,herzog2009lectures}. Through the AdS/CFT duality, one can then gain information about the properties of the superfluid in the dual field theory by performing specific calculations on the black hole horizon in the bulk \cite{herzog2011transport}. 

Surprisingly, it seems that the present literature on holographic superfluids have yet to address the emergent spacetime dynamics of a superfluid's sound modes. A widely known result from the analog gravity program is that sound modes within fluids behave like massless scalar fields in an effective curved spacetime \cite{novello2002artificial,barcelo2011analogue}. With a careful tuning of the fluid's properties, an acoustic black hole can then be made to form within the fluid. Like its gravitational counterpart, the salient feature of an acoustic black hole spacetime is the existence of a causal boundary, here called the acoustic horizon -- a trapping outer surface that no outgoing
null particles can escape \cite{barcelo2011analogue,booth2005black}, with sound
modes taking the role of null particles instead of photons \cite{novello2002artificial,barcelo2011analogue}. More than just a theoretical
model, acoustic black holes have been produced in various experimental implementations, mostly for the purpose of laboratory simulations of Hawking radiation, black hole superradiance, and earth-bound searches for quantum gravity signatures \cite{steinhauer2016observation, vocke2018rotating,Weinfurtner2024,braunstein2023analogue}. Recent theoretical studies explored the case of acoustic black holes in fluids embedded in curved spacetime \cite{ge2019acoustic,guo2020acoustic,molla2023gravitational,wang2022scrambling}, and demonstrated that in general, information about the physical background spacetime must be encoded in the effective metric of acoustic gravity \cite{ge2019acoustic, guo2020acoustic}.

The important takeaway from this short detour into analog gravity is that the flow of the scalar field in the bulk modifies the effective metric for the sound modes. When studying the properties of the sound modes in a flowing superfluid, one must therefore use the effective metric, which is in general different from the background spacetime metric of the original scalar field. An interesting case is when the effective metric has an acoustic horizon. To our knowledge, such bulk acoustic horizons arising from the superfluid flow have never been studied in the context of AdS/CFT duality.

In this paper, we study the sound modes in an acoustic black hole geometry realized using a scalar fluid flowing in an AdS background. We want to know the effects of having a bulk acoustic horizon, in contrast to a real horizon, to the dual field theory. The rest of the paper is organized as follows: in Section \ref{sec:acousspacetime}, we derive the metric for an acoustic spacetime from the Klein-Gordon equation in a pure AdS background, and establish a criteria concerning the fluid's velocity profile so that an acoustic black hole can be admitted into the fluid. In Section \ref{sec:acousticKG}, we consider two examples and solve the acoustic Klein-Gordon equation describing the propagation of sound modes within the fluid. We determine the asymptotic behavior of the solution as $r\rightarrow\infty$, and use this to extract the source, operator expectation value, Green's function, and spectral density for the dual field theory. We show in the second example that the acoustic horizon can affect these quantities at the boundary. In Section \ref{sec:hawkingtemperature} and \ref{sec:nearh}, we calculate the effective Hawking temperature of the boundary field theory due to the presence of an acoustic horizon and investigate the possibility of an acoustic horizon to have an emergent quantum criticality. Lastly, we give our conclusions and outlook in Section \ref{sec:conclusions}.

\section{Effective metric and acoustic horizon}
\label{sec:acousspacetime}
We start by deriving the equation for the sound modes and show that this is the massless Klein-Gordon equation in an effective metric. We then explicitly derive the explicit and diagonal form for this effective metric.

Our starting point is the Klein-Gordon equation in curved spacetime
\begin{equation}\label{Eqmo}
    \Box \Psi + m^2 \Psi -b \vert \Psi \vert^2 \Psi =0.
\end{equation}
Here $\Psi$ is a complex scalar field, $b$ is the interaction strength, and $m$ is the mass of the scalar field. The effects of the background spacetime with metric $g_{\mu\nu}$ are encoded in the first term 
\begin{equation}
    \Box \Psi = \frac{1}{\sqrt{-g}}\partial_{\mu}\Bigl[\sqrt{-g}g^{\mu\nu}\partial_{\nu}\left(\sqrt{\rho}e^{i\theta} \right) \Bigr],
\end{equation}
where $g$ is the determinant of the background metric and we have written the scalar field in the form $\Psi = \sqrt{\rho}e^{i\theta}$. Equation \eqref{Eqmo} can then be split into real and imaginary parts,
\begin{equation}\label{realpart}
    \frac{\Box \sqrt{\rho}}{\sqrt{\rho}}   -g^{\mu\nu}  \partial_{\mu} \theta\partial_{\nu} \theta + m^2 -b\rho = 0
\end{equation}
and
\begin{equation}\label{impart}
    \frac{1}{\sqrt{-g}} \partial_{\mu}\Bigl( \sqrt{-g}g^{\mu\nu}\rho \partial_{\nu}\theta \Bigr) = 0, 
\end{equation}
respectively.

In the real part, the term $\frac{\Box\sqrt{\rho}}{\sqrt{\rho}}$ is
called the relativistic quantum potential, and is usually assumed negligible in the hydrodynamic approximation since acoustic gravity is concerned only with perturbations up to linear order \cite{barcelo2011analogue,RevModPhys.71.463}. We assume that the modulus and the phase can be separated into a sum of zeroth order solution and fluctuations above the zeroth order
\begin{align}
    \rho =& \rho_0 +\varepsilon \rho_1\\
    \theta =& \theta_0 + \varepsilon \theta_1,
\end{align}
where $\varepsilon$ is some small parameter to track the perturbative expansion. Expanding Eq. \eqref{realpart} and retaining only up to linear order in $\varepsilon$, we then find the zeroth order equation to be
\begin{equation}\label{zeroth_order}
    b\rho_0 = m^2 - g^{\mu\nu}\partial_{\mu}\theta_0 \partial_{\nu}\theta_0.
\end{equation}
This is the equation obeyed by the background fluid, where $\rho_0$ is the fluid's density profile and $v_{j} = \partial_{j}\theta_0$ is its velocity along the $j$th spatial direction. The speed of sound in the fluid is $c_s^2 = b\rho_0/2$. The linear order equation in its raw form reads 
\begin{equation}\label{notyetKG}
    \partial_{\mu}\Bigl[ \sqrt{-g}\Bigl(g^{\mu\nu}c^2_s - g^{\mu\alpha}v_{\alpha}v^{\nu}\Bigr) \partial_{\nu}\theta_1 \Bigr]=0.
\end{equation}
This can be cast into the form of a Klein-Gordon equation if we can rewrite
\begin{align}
     \sqrt{-g}\big(g^{\mu\nu}c^{2}_s -g^{\mu\alpha}v_{\alpha}v^{\nu} \big) = \sqrt{-\mathcal{G}}\mathcal{G}^{\mu\nu},
\end{align}
where $\mathcal{G}_{\mu\nu}$ is the effective metric tensor for the acoustic spacetime and $\mathcal{G}=\text{Det}(\mathcal{G}^{\mu\nu})$.

Since we are eventually interested in the AdS/CFT dual, we set the background to be pure anti-de Sitter space for simplicity:
\begin{equation}\label{AdSmetric}
    ds^2 = -f(r)dt^2 + \dfrac{1}{f(r)}dr^2 + dS_2^2 ,
\end{equation}
 where $f(r) = 1+ r^2/L^2 $, $L$ being the AdS radius and the cosmological constant is $\Lambda = -3/L^2$. 
 
For a fluid flowing only through the radial direction,
$\mathcal{G}^{\mu\nu}$ then takes the explicit form
\begin{equation}\label{firstGmatrixup}
    \mathcal{G}^{\mu\nu} = \frac{1}{\sqrt{-\mathcal{G}}}   \begin{pNiceMatrix}
        -\frac{1}{f}\left(c^2_s +\frac{1}{f}v^2_t \right) & v_tv_r & 0 & 0 \\
       v_tv_r & f\left(c^2_s -fv^2_r \right) & 0 & 0 \\
        0 & 0 & c^2_s & 0 \\
        0 & 0 & 0 & c^2_s 
    \end{pNiceMatrix}.
\end{equation}
Inverting this then gives us the metric for the effective acoustic spacetime as
\begin{align}
\label{metric2}
    ds^2 =& \frac{\sqrt{-\mathcal{G}}}{c_s^2} \Bigg[\frac{c_s^2}{A} \Bigg(f(r)\left(c^2_s -v_{r}v^{r} \right)dt^2 - 2v_tv_rdrdt\nonumber\\
    &-\frac{1}{f(r)} \left(c^2_s -v_{t}v^{t} \right)dr^2 \Bigg)+dS_2^2\Bigg].
\end{align}
where 
\begin{align}
   A &= c^2_s\left(-c^2_s +fv^2_r -\frac{v^2_t}{f} \right) = - c^2_s\left(c^2_s -v_{\mu}v^{\mu} \right)
\end{align}
and
\begin{align}
   \mathcal{G} &= -c^6_s\left(c^2_s -v_{\mu}v^{\mu} \right).
\end{align}

The presence of $f(r)$ in the metric shows that information of the background geometry is encoded into it as well. This will become important later when we investigate the acoustic horizon.

With $\mathcal{G}^{\mu\nu}$ given by Eq. \eqref{firstGmatrixup}, the linear order Eq. \eqref{notyetKG} can then be written in the form
\begin{equation}\label{acouKG}
    \dfrac{1}{\sqrt{-\mathcal{G}}}\partial_{\mu}\left( \sqrt{-\mathcal{G}} \mathcal{G}^{\mu\nu}\partial_{\nu}\theta_1\right) = 0.
\end{equation}
This is a massless Klein-Gordon equation in an effective metric Eq. \eqref{firstGmatrixup}. The propagation of sound waves in the fluid is therefore described by an effective acosutic metric that depends on the superfluid flow and is, in general, different from the background spacetime metric.

Let us simplify the effective metric Eq. \eqref{metric2} further, and convert it into diagonal form to make the later calculations convenient. To do this, first observe that the fluid's four velocity satisfies
\begin{align}
    g^{\mu\nu}v_{\mu}v_{\nu}=v^{\mu}v_{\mu}=-1,
\end{align}
so that 
\begin{align}
    c_s^2 - v_{\mu}v^{\mu}= c_s^2+1
\end{align}
and thus $\mathcal{G}$ and $A$ are both negative. Re-writing $A = -\vert A \vert$ and transforming $dt \rightarrow \sqrt{\vert A \vert}dt$, $dr \rightarrow \sqrt{\vert A \vert}dr$, Eq. \eqref{metric2} can be recasted as 
\begin{align}
        ds^2 &= \frac{\sqrt{-\mathcal{G}}}{c_s^2} \Bigg[c_s^2 \Bigg(-f(r)\left(c^2_s -v_{r}v^{r} \right)dt^2 + 2v_tv_rdrdt \nonumber \\ &+\frac{1}{f} \left(c^2_s -v_{t}v^{t} \right)dr^2 \Bigg)+r^2dS_2^2\Bigg].
\end{align}

Following the previous works on acoustic black holes in curved spacetime \cite{ge2019acoustic,guo2020acoustic,vieira2021quasibound}, we can further simplify the expressions in our metric by rescaling the velocities: $ c_s \rightarrow c_s/c_s, \, v_r \rightarrow v_r/c_s, \, v_t \rightarrow v_t/c_s $. Furthermore, since the fluid's four velocity satisfies the normalization $v^{\mu}v_{\mu}=-1$ and assuming that the speed of sound $c_s$ is constant, the factor $\sqrt{-\mathcal{G}}$ just contributes to an overall constant, which can be dropped from the metric.  From this we get
\begin{align}
\label{metric}
    ds^2 &=  -f(r)\left(c^2_s -v_{r}v^{r} \right)dt^2 + 2v_tv_rdrdt \nonumber\\ &+\frac{1}{f} \left(c^2_s -v_{t}v^{t} \right)dr^2 +r^2dS_2^2.
\end{align}
    
Finally, we perform another coordinate transformation
\begin{align}
   dt \rightarrow dt+ \dfrac{v_tv_r}{f(r)\left(c^2_s -v_{r}v^{r} \right)} dr  
\end{align}
on Eq. \eqref{metric} to make it diagonal
\begin{align}
\label{conformalmetric1}
    ds^2 =& -f(r)\left(c^2_s -v_{r}v^{r} \right)dt^2 +\dfrac{c_s^2 \left(c^2_s +1 \right)}{f(r)\left(c^2_s -v_{r}v^{r} \right) }dr^2\nonumber\\
    &+r^2dS_2^2.
\end{align}

\begin{figure*}[t]
\includegraphics[width=\textwidth]{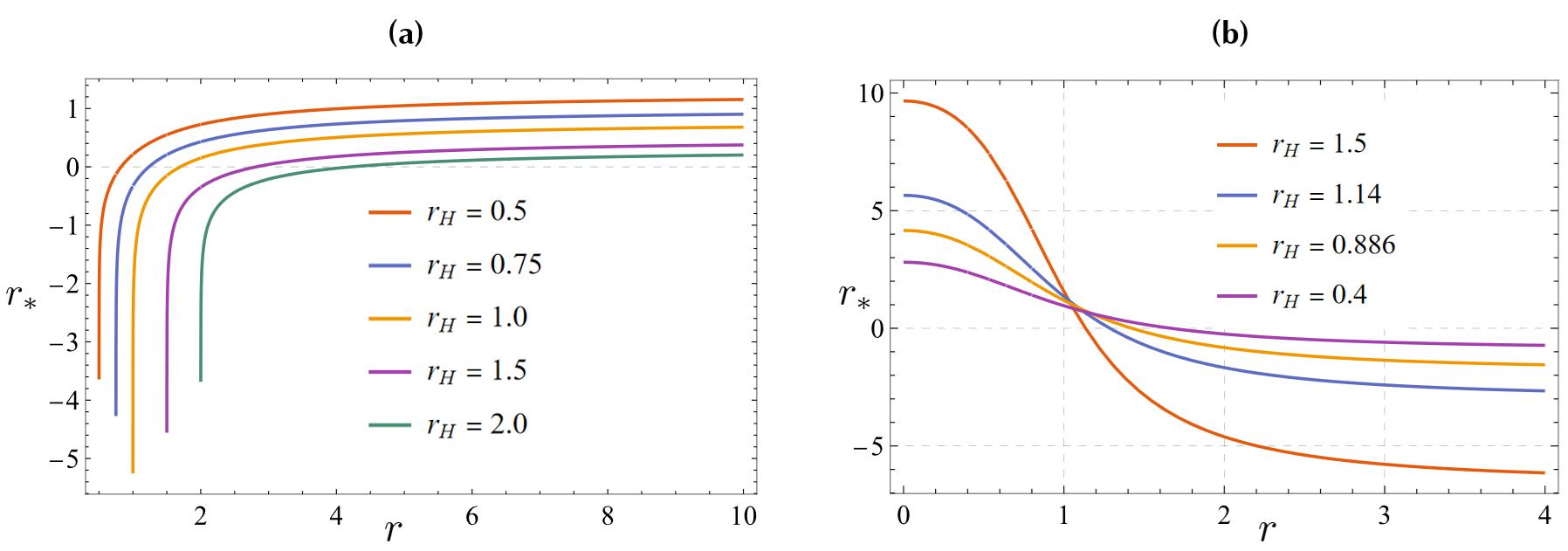}
 \caption{Behavior of the tortoise coordinates $r_*$ as a function of $r$, for different values of the acoustic horizon $r_H$, with AdS radius set to $L=1$. Figures (a) and (b) correspond to the acoustic black holes with acoustic emblackening factor $a_1(r)$ and $a_2(r)$ respectively.} 
  \label{fig:tortoise}
\end{figure*}

We will use this simple and diagonal form of the effective metric in our subsequent calculations below. This form also makes the analysis of the presence of the acoustic horizon easier. 

The guarantee for an acoustic black hole spacetime is the presence of an acoustic horizon. Sound waves in fluids get dragged along the direction of fluid flow, so in regions with supersonic flow, the upstream propagation of sound will not be able to get past a certain boundary. The supersonic region can be thought of as the interior region of the acoustic black hole and the boundary is the acoustic horizon. Any sound wave originating from the acoustic black hole interior cannot propagate upstream past the acoustic horizon. This is reflected in the function 
\begin{align}
    \label{a}
    a(r) \equiv c_s^2 - f(r)v^2_r(r) 
\end{align}
that appears in the metric Eq. \eqref{conformalmetric1}. Unlike in the flat Minkowski spacetime background, however, we do not simply compare the speed of sound $c_s$ with the superfluid speed $v_r(r)$, since Eq. \eqref{a} also depends on the AdS factor $f(r)$. Instead, we must compare $c_s$ with $v_r(r)\sqrt{f(r)}$.

We can interpret Eq. \eqref{a} as an acoustic emblackening factor since its behavior can be used as a determining characteristic of an acoustic black hole spacetime. An acoustic black hole exists if there is a radius $r_H$, called the acoustic horizon, such that
\begin{equation}
\label{conventionalhorizon}
    \begin{cases}
       a(r) <0,\;\text{for}\; r<r_H\\
       a(r) >0,\;\text{for}\; r>r_H.
    \end{cases}
\end{equation}

To give a simple example, one such function $a(r)$ is 
\begin{align}\label{a1}
    a_1(r) = c^2_s-c^2_s\frac{r_H}{r}.
\end{align}
 This corresponds to the velocity profile
\begin{align}
\label{example1velocity}
   v(r) = -c_s L \sqrt{\frac{r_H}{r(r^2+L^2)}}
\end{align}
which describes a radially inward flow. The superfluid velocity vanishes at the boundary $r\rightarrow\infty$. 

In the next sections, we will use Eq. \eqref{a1} as one of our examples to illustrate the holographic dual and the physics of the acoustic horizon.

\section{Source, expectation value, and green's function}
\label{sec:acousticKG}
We have seen that the sound waves are massless scalar waves in the acoustic spacetime and obey Eq. \eqref{acouKG} with the metric Eq. \eqref{conformalmetric1}. In the following, we now assume the existence of an acoustic horizon, realized by some suitable radial fluid flow.
Our equation for the propagation of radial sound modes then become
\begin{equation}\label{eomKG}
   \Tilde{\mathcal{G}}^{tt}\partial^2_t \theta_1(r,t) + \partial_r \Tilde{\mathcal{G}}^{rr} \partial_r\theta_1(r,t) + \Tilde{\mathcal{G}}^{rr}\partial^2_r \theta_1(r,t) = 0.
\end{equation}

After a Fourier transformation in time, we obtain the differential equation
\begin{equation}\label{KGODE}
    \dfrac{d}{dr}\Bigg[ f(r)a(r)\dfrac{d}{dr}\Tilde{\theta}(r,\omega) \Bigg] +\dfrac{\omega^2}{ f(r)a(r)}\Tilde{\theta}(r,\omega) = 0, 
\end{equation}
where the constant $c_s^2 \left(c^2_s +1 \right)$ has been absorbed into the parameter $\omega$. 

We now consider two specific examples that are motivated by simplicity and experiments.

\subsection{Acoustic Black Hole 1}
For our first example, we consider the emblackening factor given by Eq. \eqref{a1}. The exact solution to Eq. (\ref{KGODE}) is
\begin{align}\label{soln}
    \Tilde{\theta}(r,\omega) &= D_1 \cos{\omega r_{*}} + D_2 \sin{\omega r_{*}}\nonumber\\
    &=\frac{1}{2}(D_1-iD_2)e^{i\omega r_*}+\frac{1}{2}(D_1+iD_2)e^{-i\omega r_*},
\end{align}
where $D_1,D_2$ are constants and $r_{*}$ is the tortoise coordinate  
\begin{align}\label{tortoise}
    r_{*} =& \dfrac{1}{2} \left( 1+\frac{r^2_H}{L^2}\right)^{-1}\bigg[ 2L\tan^{-1}(r/L) +2r_H\ln{\left(r-r_H \right)}\nonumber\\
    &-r_H\ln{\left(r^2+L^2\right)}\bigg].
\end{align}

In the second line of Eq. \eqref{soln}, combining with the time evolution $e^{-iEt}$, we see that the two terms describe outgoing and incoming sound propagation, respectively. Given the behavior of the tortoise-like coordinate $r_*$ as a function of $r$, as shown in Fig.\ref{fig:tortoise} (a), we are able to deduce that purely ingoing sound modes are then $\Tilde{\theta}_{in}(r,\omega) \sim  e^{i\omega r_*} $. 

 Near the AdS boundary ($r\rightarrow \infty$), we can solve for the asymptotic behavior of $\Tilde{\theta}(r,\omega)$  by expanding it as a series in powers of $\dfrac{1}{r}$. To account for the plane wave propagation of the sound modes with wave vector $\mathbf{k}$ orthogonal to the radial direction, we simply replace $\omega^2 \mapsto \omega^2-k^2 $. The asymptotic solution for ingoing sound modes then take the form

 \begin{widetext}
\begin{align}
    \Tilde{\theta}_{\text{in}}(r,\omega, \mathbf{k}) &\sim a_{-}\Bigg\{ \cos{\Bigg(\dfrac{\pi L\sqrt{\omega^2 - \mathbf{k}^2}}{2f(r_H)} \Bigg)} + i\sin{\Bigg(\dfrac{\pi L\sqrt{\omega^2 - \mathbf{k}^2}}{2f(r_H)} \Bigg)}\nonumber \\ &+L\sqrt{ \omega^2 - \mathbf{k}^2} \Bigg[ \sin{\Bigg(\dfrac{\pi L\sqrt{\omega^2 - \mathbf{k}^2}}{2f(r_H)} \Bigg)} -i \cos{\Bigg(\dfrac{\pi L\sqrt{\omega^2 - \mathbf{k}^2}}{2f(r_H)} \Bigg)}\Bigg] \Bigg(\dfrac{r}{L}\Bigg)^{-1} + \ldots \Bigg\}.
\end{align}
\end{widetext}

From the ADS/CFT correspondence \cite{Gubser1998, Witten1998}, we map the coefficients of the leading and subleading terms in the asymptotic expansion to the source,

\begin{equation}\label{source_timelikeingoing}
    J_{\text{in}}(\omega, \mathbf{k)} \sim  \exp\{i\phi(\omega,\mathbf{k},r_H)\},
\end{equation}
and to the operator expectation value,
\begin{align}\label{OVEV_timelikeingoing}
    \langle \mathcal{O}_{\text{in}} (\omega, \mathbf{k})\rangle &\sim  -iL\sqrt{ \omega^2 - \mathbf{k}^2} \exp\{i\phi(\omega,\mathbf{k},r_H)\},
\end{align}
respectively, where 
\begin{align}
\label{phasexample1}
    \phi(\omega, \mathbf{k},r_H)\equiv \dfrac{\pi L\sqrt{\omega^2 - \mathbf{k}^2}}{2f(r_H)}.
\end{align}

We can see that for time-like momenta, $\omega>|\mathbf{k}|$, both the source and operator expectation value are oscillatory. For space-like momenta, $\omega<|\mathbf{k}|$, the phase Eq. \eqref{phasexample1} becomes imaginary and, consequently, the source and operator expectation value are either exponentially growing or decaying. Regularity demands that we must choose the exponentially decaying solution.

The retarded Green's function is then obtained as the ratio of the operator expectation value and the source from the solution of purely ingoing sound modes. In this case,
\begin{align}\label{retardedGreens_timelike}
    G_{\text{ret}}(\omega, \mathbf{k})  = -i  L\sqrt{ \omega^2 - \mathbf{k}^2}.
\end{align}

The spectral density function is then easily obtained as
\begin{align}
    \mathcal{A}(\omega, \mathbf{k}) =& - \text{Im} \,G_{\text{ret}}(\omega, \mathbf{k})\\
    =&
    \begin{cases}
        L\sqrt{ \omega^2 - \mathbf{k}^2},\;\;&\omega\geq|\mathbf{k}|\\
        0,\;\;&\omega<|\mathbf{k}|.
    \end{cases}
\end{align}
Note that for this example, the dependence on the horizon $r_H$ dropped in the final expression for the Green's function and spectral density. From Eqs. \eqref{source_timelikeingoing} and \eqref{OVEV_timelikeingoing}, we see that the horizon enters only through the phase Eq. \eqref{phasexample1}. The source Eq. \eqref{source_timelikeingoing} and operator expectation value Eq. \eqref{OVEV_timelikeingoing} are in phase which leads to the cancelation of the horizon dependence.

\begin{widetext}

\begin{figure}
    \centering
    \includegraphics[width=\textwidth]{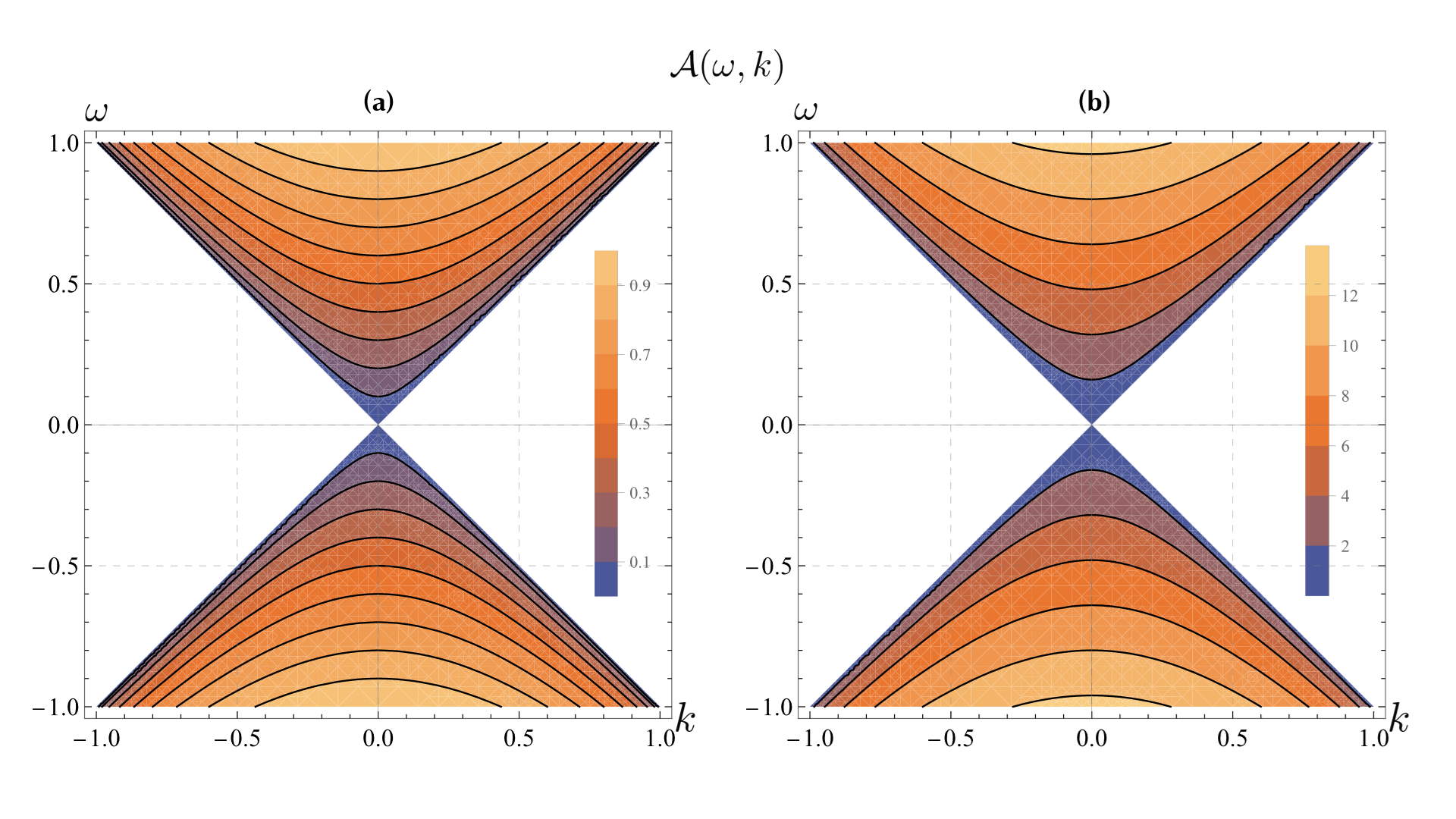}
    \caption{Spectral density for the dual field theory at the boundary for the case of AdS acoustic black hole spacetimes characterized by the acoustic emblackening factors $a_1(r)$ (a) and $a_2(r)$ (b). The AdS radius has been set to $L=1$. For (b), the plot is generated with $Z=0.04$.} 
    \label{fig:CFT results}
\end{figure}

\end{widetext}

The plot of the spectral density is shown on Fig. \ref{fig:CFT results} (a). We can see the typical gapless profile of the sound modes, which is consistent with the fact that these are Goldstone modes of the broken U(1) symmetry. Instead of a well-defined cone that gives poles to the Green's function, however, we see branch cuts, which is typical of strongly-coupled dual field theories.


\subsection{Acoustic Black Hole 2}
For our next example, we consider the velocity profile \begin{equation}\label{1sqrtr}
    v_r(r) = -\sqrt{\dfrac{Z}{r} },
\end{equation}
where $Z$ is a constant. This describes an inwardly flowing fluid that goes faster for decreasing $r$. We therefore expect an acoustic horizon to form at some radius, which we solve explicitly below. This velocity ansatz has been used in many studies of acoustic black holes including laboratory simulations of black hole physics \cite{novello2002artificial, PhysRevA.78.063804, torres2017rotational, guo2020acoustic, vocke2018rotating}. In ref. \cite{guo2020acoustic}, the physical motivation for this ansatz is provided by arguing that in the presence of a massive compact object, the escape velocity $v_e$ of an observer that remains stationary at a Schwarzschild coordinate radius $r$ takes the form $v_e \sim \sqrt{\dfrac{2M}{r}}$.  

The acoustic emblackening factor that corresponds to the velocity profile in Eq. \eqref{1sqrtr} is 
\begin{equation}\label{2ndblackfactor}
    a_2(r) = c_s^2 - \left(1+\dfrac{r^2}{L^2}\right) \dfrac{Z}{r}.
\end{equation}

\begin{figure}[h!]
\includegraphics[width=0.48\textwidth]{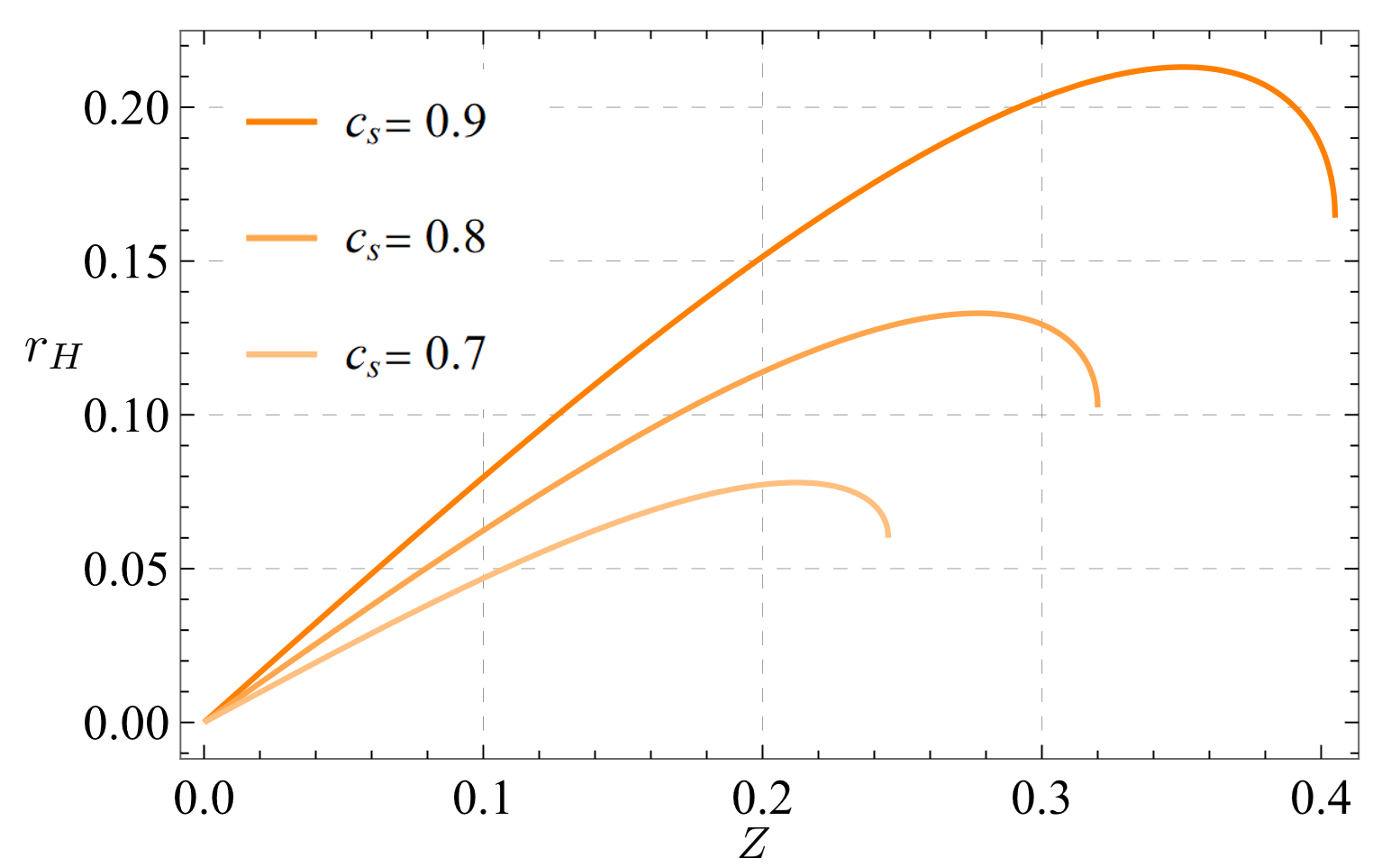}
 \caption{$Z-$ dependence of the horizon radius  associated with the acoustic emblackening factor $a_2(r)$ (Eq. \eqref{2ndblackfactor}), plotted for different values of the sound speed $c_s$. Here, AdS radius has been set to $L=1$.} 
  \label{fig:rZ}
\end{figure}

In this case, the acoustic horizon radius is
\begin{equation}\label{horizon2}
    r_H = \dfrac{c_s^2L^2}{2Z}\left[1 + \sqrt{1-\left(\dfrac{2Z}{c_s^2L} \right)^2}\right].
\end{equation}
From the expression above, it is easy to read off the constraints on the values of $Z$ for the horizon radius to be physical:
\begin{equation}\label{Zconstraint}
    0 < Z  \leq \dfrac{c_s^2L}{2}.
\end{equation}
In Fig. \ref{fig:rZ}, we plot the dependence of the horizon radius $r_H$ on the parameter $Z$ for different values of the speed of sound, $c_s$. The plots show that $r_H$ has an upper bound. Note that the speed of sound also has an upper limit -- it cannot be equal to 1. Based on Fig. \ref{fig:rZ}, we can safely infer that $r_H < L$, which means that the acoustic black holes produced by the flowing fluid with velocity Eq. \eqref{1sqrtr} will all be smaller than the AdS radius.

Solving the acoustic Klein-Gordon equation \eqref{eomKG}, we then find the asymptotic behavior of ingoing sound modes for this case to be 

\begin{equation}\label{ABHSingosoun}
    \Tilde{\theta}_{\text{in}}(r,\omega, \mathbf{k}) \sim a_{-} e^{-i\sqrt{\omega^2-\mathbf{k}^2} \chi} \left[1- \dfrac{i\sqrt{\omega^2-\mathbf{k}^2}L^2}{2Z} \left(\dfrac{r}{L}\right)^{-2} \right] 
\end{equation}
where 
\begin{equation}\label{rtilde}
    \chi = \dfrac{\pi L}{2c^2_s} \left(1 - 2i\sqrt{\dfrac{Z}{c_s^2 - 4Z^2}} \right).
\end{equation}

Equation \eqref{ABHSingosoun} gives us the source and the operator expectation values as the leading and sub-leading terms, respectively. The retarded Green's function can then be obtained using the AdS/CFT correspondence, 

\begin{align}\label{retardedGreens2}
    G_{\text{ret}}(\omega, \mathbf{k})  = -i  \frac{L^2}{2Z}\sqrt{ \omega^2 - \mathbf{k}^2},
\end{align}
 which leads to the following spectral density function
\begin{equation}\label{spec2}
    \mathcal{A}(\omega, \mathbf{k}) = 
    \begin{cases}
        \dfrac{\sqrt{\omega^2-\mathbf{k}^2} }{2Z},\;\;&\omega\geq|\mathbf{k}|\\
        0,\;\;&\omega<|\mathbf{k}|.
    \end{cases}
\end{equation}
Observe that we get the same $\sqrt{\omega^2-\mathbf{k}^2}$ dependence. However, this spectral density now depends on the horizon via the parameter $Z$ given by Eq. \eqref{horizon2}.

\section{Effective Hawking Temperature}
\label{sec:hawkingtemperature}
It is well known that acoustic black holes also emit Hawking radiation and possess an effective Hawking temperature \cite{Unruh1981}. This was confirmed by various experiments using atomic Bose-Einstein condensate \cite{Steinhauer2016, Muñoz2019}, optics \cite{Belgiorno2010, Steinhauer2014, Drori2019}, polariton superfluids \cite{Nguyen2015}, and even ordinary fluids \cite{Weinfurtner2011, Euv2016}.

The presence of an acoustic horizon suggests that the sound modes of the dual field theory have a different effective temperature compared to the original scalar field since the former is governed by the effective metric Eq. \eqref{conformalmetric1}, while the latter is governed by the background spacetime metric. The effective temperature in the dual theory should be given by the Hawking temperature of the acoustic horizon.

We now investigate this effective Hawking temperature $T_H$ for the sound modes. Since our effective metric for an AdS acoustic black hole Eq. \eqref{conformalmetric1} is similar in form to that of a general static black hole, we may compute for $T_H$ by following the Euclidean-gravity insight from Gibbons and Hawking, which is also illustrated in Reference \cite{zaanen2015holographic}. 

\begin{figure*}[t]
\includegraphics[width=\textwidth]{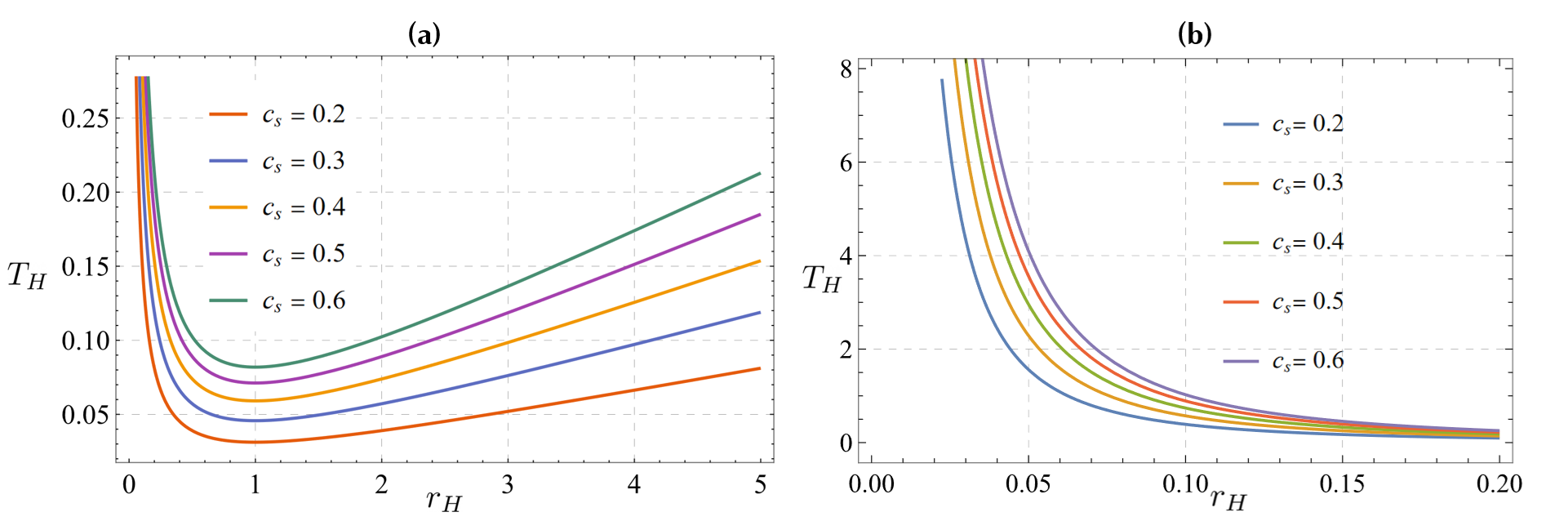}
  \caption{Effective Hawking temperature of acoustic black holes defined by acoustic emblackening factor $a_1(r)$ (a) and $a_2(r)$ (b) respectively. Portrayed is the temperature behavior as a function of the acoustic horizon radius for different values of the sound speed in the fluid $c_s$ and with $Z=0.25c^2_s$. }
  \label{fig:TempH}
\end{figure*}

First, we write the metric Eq. \eqref{conformalmetric1} in the form
\begin{equation}
    ds^2 = -h(r)dt^2 + \dfrac{dr^2}{h(r)} + r^2 dS^2_2,
\end{equation}
 identifying $h(r)=f(r)a(r)$ to be a general emblackening factor. We call it general in the sense that it is a product of $f(r)$ and $a(r)$, which vanish at the gravitational and acoustic horizons respectively, whenever the spacetime admits both a real and an acoustic analog black hole.  Since in this paper, we are operating on a pure AdS background, there is no gravitational black hole. Focusing on the region near the acoustic horizon, we expand the metric about $r=r_H$ and keep only the leading term, then Wick rotate to Euclidean time $\tau =it$. We get
\begin{align}
    ds^2= h'(r_H) \qty(r-r_H) d\tau^2
    +\frac{c_s^2(c_s^2+1)}{h'(r_H) \qty(r-r_H)}dr^2
    + r^2 dS^2_{2}
\end{align}
where $h'(r_H)\equiv f(r_H)a'(r_H)$ and a primed $h'(r_H)$ and  $a'(r_H)$ means derivative with respect to the radius and evaluated at the horizon.

Changing coordinates along the radial direction,
\begin{align}
    R_0=2\frac{\sqrt{c^2_s\qty( c_s^2+1)}\sqrt{r-r_H}}{\sqrt{h'(r_H)}}
\end{align}
we get 
\begin{equation}
    ds^2 =\dfrac{1}{4} \dfrac{\qty(h'(r_H))^2 }{c^2_s\qty(c_s^2+1)} R^2_0 d\tau^2 +dR^2_0 +r^2dS^2_{2}.
\end{equation}
Making another change of coordinates in Euclidean time,
\begin{align}
    \Theta=\dfrac{1}{  2}  \dfrac{h'(r_H)}{ \sqrt{c^2_s\qty(c_s^2+1)}  } \tau, 
\end{align}
the near-horizon metric becomes
\begin{align}\label{eucperiodic}
    ds^2=R_0^2d\Theta^2 + dR_0^2 + r^2dS^2_{2}.
\end{align}

Observe that the $R_0-\Theta$ part of Eq. \eqref{eucperiodic}  is just the polar coordinates of a plane and must, therefore, obey the periodic boundary condition on $\Theta+2\pi=\Theta$. This implies that the Euclidean time $\tau$ has the periodicity
\begin{align}
\label{period}
    P_{\tau}=\frac{4\pi\sqrt{c^2_s(c_s^2+1)}}{h'(r_H)}.
\end{align}

The reciprocal of Eq. \eqref{period} gives us the effective Hawking temperature of the acoustic horizon
\begin{align}
\label{effectivetemperature}
    T_H=\frac{h'(r_H)}{4\pi\sqrt{c^2_s(c_s^2+1)}}.
\end{align}

For the AdS acoustic black hole of our first example, which is characterized by the acoustic blackening factor in Eq.\eqref{a1}, the effective Hawking temperature is
\begin{align}
\label{effectivetemperature1}
    T_H = \dfrac{c_s}{4\pi \sqrt{c_s^2+1}} \qty(r_H + \dfrac{1}{r_H}).
\end{align}

In Fig. \ref{fig:TempH} (a), we show how this Hawking temperature varies with the horizon radius. For acoustic black holes significantly smaller than the AdS radius, the effective Hawking temperature varies inversely with the horizon radius $T_H \sim 1/r_H$, which similar to the asymptoticaly flat Schwarzschild black hole. But this similarity is soon lost once the acoustic horizon radius becomes comparable in size with the AdS radius. As the acoustic horizon becomes larger than the AdS radius, the $r_H$-dependence of its Hawking temperature transitions to a linear relation $\qty(T \sim r_H)$, similar to that found in the case of Schwarzschild-AdS black holes \cite{zaanen2015holographic}. 

For the second acoustic black hole, characterized by the acoustic emblackening factor in Eq. \eqref{2ndblackfactor}, the first derivative of the general emblackening factor is 
\begin{equation}\label{h2prime}
    h'(r_H) = -Z \qty(1+r^2_H)\qty(1-\frac{1}{r^2_H}),
\end{equation}
giving us an effective Hawking temperature of 

\begin{equation}\label{THtwo}
    T_H = \frac{-Z}{4\pi\sqrt{c^2_s(c_s^2+1)}}  \qty(r^2_H -\frac{1}{r^2_H}),
\end{equation}
where the constant $Z$ is a positive number with upper bound as specified by Eq. \eqref{Zconstraint}. From Eq. \eqref{Zconstraint} and Fig. \ref{fig:rZ}, we saw that the acoustic horizon is bounded $r_H<1$. The temperature above is therefore nonnegative. We show the dependence of the effective Hawking temperature with the horizon radius in Fig. \ref{fig:TempH} (b). It approaches zero as $r_H$ approaches the AdS radius $L=1$.

\section{Near-horizon effective geometry}
\label{sec:nearh}
The extremal Reissner–Nordström black holes in AdS spacetime is known to give rise to an emergent quantum criticality in the infrared region \cite{Faulkner2011}. This comes from the fact that the near-horizon expansion of the emblackening factor gives doubles zeros. That is, the leading order in the expansion about $r=r_H$ is the second order $(r-r_H)^2$. As a consequence, the geometry near the horizon is AdS$_2\times\mathbb{R}^{d-1}$. This is interpreted as a local quantum criticality as can be seen from the resulting scaling symmetry $t\rightarrow\lambda^zt$ and $\mathbf{x}\rightarrow\lambda\mathbf{x}$ with $z\rightarrow\infty$ \cite{Faulkner2011, zaanen2015holographic}.

We now investigate if this can occur in acoustic black holes. From Eq. \eqref{conformalmetric1}, the analog of the emblackening factor in our acoustic black hole is

\begin{equation}\label{emblack}
    h(r) \equiv f(r)\left(c^2_s -v_{r}v^{r} \right).
\end{equation}

Our first example Eq. \eqref{example1velocity} yields the expansion for the emblackening factor
\begin{align}
    h(r)\approx\frac{c_s^2}{r_H}(1+r_H^2)(r-r_H)+\frac{c_s^2}{r_H^2}(r_H^2-1)(r-r_H)^2,
\end{align}
while the second example Eq. \eqref{1sqrtr} yields
\begin{align}
h(r)\approx & -\frac{c^2 }{r_H}\left(r_H^2-1\right) (r-r_H)\\
&-\frac{c^2 \left(2 r_H^4-r_H^2+1\right)}{r_H^4+r_H^2}(r-r_H)^2.\nonumber
\end{align}
Both have linear order $(r-R_H)$ as the first term in the expansion, which shows that there are no double zeros at the horizon. Hence, these examples do not have an emergent quantum criticality in the infra red region.

What kind of fluid flow produces double zeros? To investigate this, we write the emblackening factor in a general form
\begin{align}
\label{emblack2}
    h(r)=f(r)a(r).
\end{align}

An expansion about $r=r_H$ would then yield
\begin{align}
\label{nearh1}
    h(r)\approx &(1+r_H^2)a'(r_H)(r-r_H)\\
    &+\left[2r_Ha'(r_H)+\frac{1}{2}(1+r_H^2)a''(r_H)\right](r-r_H)^2,\nonumber
\end{align}
where the primes in $a'(r_H)$ and $a''(r_H)$ denote derivatives with respect to $r$ evaluated at $r=r_H$.

To get double zeros, the linear order in $(r-r_H)$ should vanish, which means
\begin{align}
\label{aprime}
    a'(r_H)=0,
\end{align}
or in terms of the superfluid velocity, using Eq. \eqref{a},
\begin{align}
\label{vprime}
    v'(r_H)=-\frac{r_Hv_r(r_H)}{1+r_H^2}.
\end{align}

The usual horizons, where the emblackening factor obeys Eq. \eqref{conventionalhorizon}, will have an inflection point at $r_H$ so that $a''(r_H)=0$. Combined with Eq. \eqref{aprime}, the second order term in Eq. \eqref{nearh1} also then vanishes and there are no double zeros.

To get double zeros, we go back to the emblackening factor, which vanishes at the horizon to yield
\begin{align}
    c_s^2-f(r_H)v_r^2(r_H)&=0\\
\label{vrh}
    v_r(r_H)&=\frac{-c_s}{\sqrt{1+r_H^2}}.
\end{align}

Substituting this into Eq. \eqref{vprime}
\begin{align}
\label{vrhprime}
    v_r'(r_H)=\frac{c_s}r_H{(1+r_H^2)^{3/2}}.
\end{align}

Equations \eqref{vrh} and \eqref{vrhprime} gives us the necessary behavior of the superfluid flow near $r_H$
\begin{align}
\label{vnearrh}
    v_r(r)\approx-\frac{c_s}{\sqrt{1+r_H^2}}+\frac{c_sr_H}{(1+r_H^2)^{3/2}}(r-r_H).
\end{align}

That is, if the background superfluid has a velocity profile given by Eq. \eqref{vnearrh} near the acoustic horizon, then the effective metric felt by the sound modes will have double zeros. Explicitly, changing the time coordinates $t\rightarrow t/(c_s^2+1)$ then expanding the emblackening factor Eq. \eqref{emblack2} about the horizon, the metric becomes
\begin{align}
    ds^2=-\frac{(r-r_H)^2}{L_2^2}dt^2+\frac{L_2^2}{(r-r_H)^2}dr^2+d\mathbf{x}^2,
\end{align}
where
\begin{align}
\label{l2}
    L_2\equiv \frac{1}{c_s}\sqrt{\frac{(r_H^2+1)(c_s^2+1)}{2r_H^2-1}}.
\end{align}

A further change of coordinates
\begin{align}
    \zeta\equiv \frac{L_2^2}{r-r_H},
\end{align}
gives
\begin{align}
    ds^2=\frac{L_2^2}{\zeta^2}(-dt^2+d\zeta^2)+d\mathbf{x}^2.
\end{align}

This shows an AdS$_2$ near-horizon geometry with effective radius given by Eq. \eqref{l2}, which is associated with quantum criticality, when the superfluid flow has the velocity profile given by Eq. \eqref{vnearrh} near $r_H$.

\section{Conclusions}
\label{sec:conclusions}
In this paper, we explored the consequences of a bulk acoustic horizon to the sound modes of the dual superfluid. We derived the effective metric for the acoustic spacetime within a scalar fluid embedded in a pure AdS background, and from the metric we identified the acoustic emblackening factor, Eq. \eqref{a}. We then use the acoustic emblackening factor to formulate the necessary conditions for an acoustic black hole to exist within the scalar fluid, cf. Eq. \eqref{conventionalhorizon}. 

We considered two superfluid velocity profiles that lead to an AdS-acoustic black hole spacetime and solved the corresponding Klein-Gordon equation, focusing on the solution that describes ingoing sound modes. After obtaining the asymptotic form of the ingoing solution near the AdS boundary, we used the GPKW rule to map the leading and sub-leading terms of the asymptotic solution to the source and operator expectation value of the dual field theory. This then lead us to the retarded Green's function and spectral density function of the dual field theory. For both cases, a gapless profile is observed for the sound modes, consistent with the fact that these are Goldstone modes of the broken $U(1)$ symmetry. Instead of a well-defined cone, the spectral density profile has several branch cuts, a feature which is typical of strongly-coupled theories. In our second example, we found that the retartded Green's function, and hence the spectral density, is affected by the location of the horizon. 

We also calculated the Hawking temperature for both AdS-acoustic black hole spacetimes. We note that the dual superfluid itself is at zero temperature, since the background metric of our original scalar field is pure anti-de Sitter. The sound modes, however, feel an effective nonzero temperature and this is due to the deformation of the original pure AdS background into an effective acoustic black hole spacetime.

In calculating the Hawking temperature at the acoustic horizon, we also obtained the near-horizon effective geometry for our AdS acoustic black hole spacetimes. Expanding the general emblackening factor about the horizon radius, we find that for both cases we considered, the leading term of the expansion is of linear order in $(r-r_H)$, thus showing that there are no double zeros at the horizon. This means that both acoustic spacetimes we considered do not have an emergent quantum criticality in the infrared region. To complete our discussion on quantum criticality, we solved for the necessary behavior of the superfluid flow near the horizon, for an acoustic black hole spacetime to possess quantum criticality.

These calculations show that the behavior of the sound modes in a flowing fluid can be different from the underlying scalar field due to the deformation of the background metric into an effective metric. In the presence of acoustic horizon, the detailed behavior depends on the specific fluid velocity profile. In our calculations above, the boundary dual is a superfluid instead of a superconductor. It would be interesting to see what happen if the sound modes are gapped out via the Anderson-Higgs mechanism. In addition, we have not explored the consequence of having a quantum criticality in the infrared region when the acoustic horizon has double zeros. We leave these interesting questions for future work.

\begin{acknowledgments}
J.C. Candare acknowledges financial support from the DOST-SEI through the ASTHRDP scholarship.
\end{acknowledgments}

\bibliography{apssamp}

\end{document}